\documentclass[aps,prd,amsmath,amsfonts,a4paper,11pt,reprint,twocolumn,square,numbers,showpacs,superscriptaddress,floatfix,sort&compress]{revtex4-1}
\usepackage{hyperref}
\usepackage{amssymb}
\usepackage{bm}
\usepackage{natbib}
\usepackage{graphicx}
\expandafter\let\csname equation*\endcsname\relax
\expandafter\let\csname endequation*\endcsname\relax
\usepackage{amsmath}
\usepackage[labelfont={bf,scriptsize},textfont={scriptsize},justification=RaggedRight,format=hang]{caption,subfig}
\usepackage{color}

\RequirePackage{ifpdf}
\ifpdf
\else 
\fi

\renewcommand{\d}{\mathrm{d}}

\newcommand{\fNL}{f_{\mathrm{NL}}}

\newcommand{\Amplitude}{A_{\mathrm{NL}}}

\newcommand{\LO}{\text{\textsc{lo}}}
\newcommand{\NLO}{\text{\textsc{nlo}}}

\newcommand{\nsq}{n_{\text{\text{sq}}}}
\newcommand{\nbias}{n_{\delta b}}

\newcommand{\db}{n}

\newcommand{\Omegam}{\Omega_{\mathrm{m},0}}
\newcommand{\halo}{\mathrm{h}}
\newcommand{\Lbias}{b_{\mathrm{L}}}
\newcommand{\deltabias}{\delta b}

\newcommand{\deltac}{\delta_{\mathrm{c}}}

\newcommand{\keq}{k_{\text{eq}}}

\newcommand{\kL}{k_{\text{\textsc{l}}}}
\newcommand{\kS}{k_{\text{\textsc{s}}}}
\newcommand{\vectkL}{\vect{k}_{\text{\textsc{l}}}}
\newcommand{\vectkS}{\vect{k}_{\text{\textsc{s}}}}

\newcommand{\B}[1]{\mathcal{B}^{{#1}}}
\renewcommand{\P}[1]{\mathcal{P}^{{#1}}}

\newcommand{\vect}[1]{\bm{\mathrm{{#1}}}}

\newcommand{\para}[1]{\par\vspace{2mm}\noindent\textbf{\emph{{#1}}.---}}

\DeclareMathOperator{\Or}{O}

\begin{document}

	\title{Scale-dependent bias from multiple-field inflation}

	\author{Mafalda Dias}
	\email{M.Dias@sussex.ac.uk}
	\affiliation{Department of Physics and Astronomy, University of Sussex,
	Brighton, BN1 9QH, UK}

	\author{Raquel H. Ribeiro}
	\email{RaquelHRibeiro@case.edu}
	\affiliation{Department of Physics, Case Western Reserve University, 
	10900 Euclid Ave, Cleveland, OH~44106, USA}
	
	\author{David Seery}
	\email{D.Seery@sussex.ac.uk}
	\affiliation{Department of Physics and Astronomy, University of Sussex,
	Brighton, BN1 9QH, UK}
	
	
	\begin{abstract}
		We provide a formula for the scaling behaviour of the
		inflationary bispectrum in the `squeezed' limit where
		one momentum becomes much smaller than the other two.
		This determines the scaling of
		the 
		halo
		bias
		at low wavenumber
        and will be an
		important observable for the next generation of
		galaxy surveys.
		Our formula allows it to be predicted for the first
		time for a generic
		inflationary model with multiple light,
		canonically-normalized
		scalar fields.
	\end{abstract}
	
	\maketitle

	For the last twenty years, our information about
	the very early universe has primarily come from the
	large-scale
	anisotropies of the cosmic microwave background
	(CMB).
	But over the next decade
	fresh
	data will come from a new generation of
	large-scale structure surveys
	such as Euclid and LSST, and complementary small-scale CMB
	observations such as PIXIE
	\cite{Amendola:2012ys, *Ivezic:2008fe, *Kogut:2011xw}.
	
	These surveys will probe a new suite of observables, requiring
	the development of techniques
	by which they can be predicted from models of the early universe.
	In this \emph{Letter} we focus on an observable 
	which will be a target for future
	surveys---the \emph{scale-dependent halo bias},
	which is sensitive to the scaling of the primordial
	bispectrum in the so-called
	squeezed limit.
	It is a discriminant of inflationary
	models with more than one active
	field~\cite{LoVerde:2007ri,*Sefusatti:2009xu,
	*Giannantonio:2011ya,*Becker:2012yr,*Becker:2012je,
	Shandera:2010ei}.
	
	\para{Scale-dependent bias}%
	We cannot observe the primordial density field itself, but only
	correlations between populations of objects which
	trace its properties.
	One such population
	are halos of mass $M$,
	which form from collapse of over-dense regions.
	The clustering of these halos is
	described by the halo--halo or matter--halo power spectra
	$P^{\halo\halo}$ and $P^{\delta\halo }$,
	where $\delta$ is the primordial density contrast.
	At lowest order the
	relationship between $\delta$ and the halo
	number density is linear 
	$P^{\delta\halo }(k) \approx \Lbias P^{\delta\delta}_R(k)$
	and therefore 
	$P^{\halo \halo}(k) \approx \Lbias^2 P^{\delta\delta}_R(k)$.
	The superscript `$R$' denotes smoothing
	over a lengthscale $R$ sufficiently
	large to enclose the halo mass $M$.
	
	Corrections to $\Lbias$ come from higher-order
	correlations of $\delta$.
	They can be expressed
	using the 
	$n$-point functions of the
	curvature perturbation $\zeta$,
	which is typically used to
	characterize predictions of inflationary models.
	The leading effect comes from the three-point
	function and
	induces a $k$-dependent shift
	$\Lbias \rightarrow \Lbias + \deltabias (k)$,
	where \cite{Sheth:2001dp, *Dalal:2007cu,
	*Slosar:2008hx, Matarrese:2008nc}
	\begin{equation}
	\begin{split}
		\deltabias(k)
		=
		\frac{M_R^{-1}(k,z) }{4\pi^2 \sigma^2_M}
		 & \Big(\frac{\deltac^2}{\sigma^2_M} -1\Big)
		\int_0^\infty
		q^2 \d q \; M_R(q,z)
		\\
		& \hspace{-4mm}
		\mbox{} \times
		\int_{-1}^{1} \d \mu \; M_R(Q,z)
		\frac{B_{\zeta}(k,q,Q)}{P_{\zeta}(k)} .
	\end{split}
	\label{eq:halo-bias}
	\end{equation}
	Here,
	$z$ is the redshift and
	$Q^2 \equiv k^2 + q^2 + 2 \mu k q$;
	the critical density for spherical collapse is $\deltac \simeq 1.69$;
	and $\sigma^2_M$ is the variance
	of the tracing population of mass-$M$ halos   \footnote{There is a correction to Eq. \eqref{eq:halo-bias} 
	    proportional to $\d / \d \ln \sigma_M$
	    of the integral, obtained in
	    Refs.~\cite{Desjacques:2011mq, Adshead:2012hs, Baumann:2012bc}.
	    This correction will be suppressed relative to the term
	    retained in~\eqref{eq:halo-bias}
	    by a term proportional to the departure of
	    $P_\zeta$ and $B_\zeta$ from scale-invariance,
	    and will therefore be small
	    when the spectrum and bispectrum are nearly scale-invariant.}.
	The kernel $M_R(q,z)$ is related to the
	linearized Poisson equation
	connecting the density contrast
	$\delta(\vect{k},z)$
	and $\zeta(\vect{k})$.
	It can be written as
	\begin{equation}
		M(k,z) = \frac{2}{5} \frac{k^2}{H_0^2}
			\frac{T(k)}{\Omegam}
			\frac{g(z)}{1+z} .
	\end{equation}
	The transfer function $T(k)$ is
	normalized to unity
	as $k \rightarrow 0$,
	and $g(z)$ accounts for
	suppression of growth
	during $\Lambda$-domination.
	The present-day matter density is
	$\Omegam$. 
	To smooth $\delta(\vect{k},z)$ on the scale $R$
	we introduce a window function
	$W_R(k)$
	and define
	$M_R(k,z) \equiv W_R(k) M(k,z)$.
	Finally, the spectrum and bispectrum
	of the curvature perturbation satisfy
	\begin{align}
		\langle
			\zeta(\vect{k}_1) \zeta(\vect{k}_2)
		\rangle
		&
		= (2\pi)^3 \delta(\vect{k}_1 + \vect{k}_2)
		P_{\zeta} \\
		\langle
			\zeta(\vect{k}_1) \zeta(\vect{k}_2) \zeta(\vect{k}_3)
		\rangle
		&
		= (2\pi)^3 \delta(\vect{k}_1 + \vect{k}_2 + \vect{k}_3)
		B_{\zeta} .
	\end{align}
	
	\para{Scaling of $\deltabias$}%
	For $k$ smaller than the wavenumber
	$\keq$ corresponding to the horizon scale at
	matter--radiation equality,
	we have $T(k) \sim 1$
	and $M(k,z)$
	proportional to
	$k^2$.
	For $k<\keq$ this sets the dominant scaling
	of $\deltabias$.
	It is the complete result
	whenever the integral
	over $B_\zeta$ does not depend on $k$,
	as for the case of the local model with constant
	amplitude $f_{\rm NL}$~\cite{Sheth:2001dp, *Dalal:2007cu,*Slosar:2008hx, Matarrese:2008nc}.
	
	In this \emph{Letter}
	we focus on
	the possibility that the integral
	in \eqref{eq:halo-bias} also
	scales nontrivially with $k$.
	The precise scaling
	depends on details of the underlying
	inflationary model,
	and may constitute a useful observable.
	Because
	$M(q)$ grows steeply with $q$
	for $q < \keq$,
	and
	the window function restricts
	the integral to wavenumbers $q \lesssim R^{-1}$,
	the dominant contribution
	will come from
	the region
	$\keq \lesssim q \lesssim R^{-1}$.
    If $k \ll \keq$ then this 
    is the `squeezed' region, where one of the
    momenta in the bispectrum is much less than
    the other two.
    Therefore, for $k \ll \keq$, the halo
    bias is a probe of the squeezed limit
    of the bispectrum.

	
	
	To study this limit we denote the `squeezed'
	momentum $\kL$. The other two momenta are taken to be
	of order $\kS$, with $\kL \ll \kS$.
   	We will see below that it has not yet been possible
	to obtain reliable predictions for the bispectrum
	in this limit,
	because the presence of multiple
	large hierarchies causes perturbation theory to break down.
	But if $\kL$ is not
	too much smaller than $\kS$,
	the bispectrum can be approximated by
	a power law
	with spectral index $\nsq$,
	\begin{equation}
		B_{\zeta} \approx \frac{\B{}_{\zeta}(\kS)}{4 \kL^3 \kS^3}
		\left( \frac{\kL}{k_t} \right)^{\nsq(\kS)}
		\left\{ 1
			+
			\Or\left( \frac{\kL^2}{\kS^2} \right)
		\right\} ,
		\label{eq:b-scaling}
	\end{equation}
	where $\B{}_\zeta(\kS)$ is an overall amplitude
	depending weakly on $\kS$ and
	$k_t = \sum_i k_i$ is the perimeter of the
	momentum triangle
	$\vect{k}_1 + \vect{k}_2 + \vect{k}_3 = 0$.
	The spectral index
	$\nsq$ measures the response of
	$B_\zeta$ to a change of $\kL$ and
    may also depend weakly on $\kS$. 
    In what follows, our aim is to provide a precise formula for the 
	spectral index $\nsq$ 
	and relate it to the scaling of $\delta b$.

    The integral over $q$
    in~\eqref{eq:halo-bias} has been
    obtained for a variety of simple
    templates
    \cite{Matarrese:2008nc, Shandera:2010ei, Afshordi:2008ru, *Verde:2009hy}.
    Here we focus on the scale dependence inherited from the
    squeezed limit of the bispectrum.
    The amplitude and scale-dependence of the
    spectrum and bispectrum can be expressed
	in terms of an arbitrary reference scale $k_\ast$.
	We write
	\begin{equation}
		P_\zeta(k) =
			\frac{\P{}_\zeta(k)}{2k^3}
			=
			\frac{\P{}_\zeta(k_\ast)}{2k^3}
			\left( \frac{k}{k_\ast} \right)^{n_s(k_\ast)-1} ,
	\end{equation}
	where
	$n_s-1\equiv \d \ln \P{}_\zeta / \d \ln k$ is the spectral index
	of $P_{\zeta}$,
	and $\B{}_\zeta(\kS) = \B{}_\zeta(k_\ast) (\kS/k_\ast)^\alpha$.
	We assume $n_s$ and $\nsq$ to be approximately constant over the
	range $\keq \lesssim q \lesssim R^{-1}$.
	Because this is comparatively small,
    with $R \keq \sim 0.1$ on cluster scales,
    we expect
    this to be reasonable.
	We conclude that $\deltabias$
	has approximate scaling behaviour
    \begin{equation}
        \deltabias \approx
            \Amplitude \times
            M_R(k,z)^{-1}
            \left(
                \frac{k}{k_\ast}
            \right)^{\nsq - (n_s - 1)} ,
        \label{eq:delta-b-scaling}
    \end{equation}
    where the 
    amplitude $\Amplitude$
    depends on details of the model,
    the scale $R$
    and the background cosmology.
    In general,
    it may be difficult to calculate. 
    For the purposes of this paper we do
    not require a precise estimate,
    because the observable is the
    scaling behaviour of $\deltabias$
    rather than its amplitude.
    This is a strength of the approach.
%

	We define the combination
	$\nbias \equiv \nsq - n_s + 1$
	as the \emph{spectral index of the bias}.
	The main result of this \emph{Letter} is
	a prescription
	to compute it for any inflationary model.
	Constraints on $\nbias$ have been obtained from
	present-day data~\cite{Giannantonio:2011ya},
	although the results show sensitivity
	to priors.
	 In particular, the first Planck results impose strong constraints on the value of $\fNL$ local~\cite{Ade:2013ydc} and therefore on the range of models that can show an observable $\nbias$. 
	Improvements are expected from a
    future Euclid- or LSST-like survey
    \cite{Sefusatti:2012ye, *Biagetti:2013sr}.
	
	The $k_t$-dependence of the bispectrum
	$B_\zeta$
	was first studied by Chen~\cite{Chen:2005fe}.
	Formulae for a multiple-field model
	were given by
	Byrnes et al. \cite{Byrnes:2010ft,Byrnes:2012sc}.
	Our analysis differs because
	of its focus on
	scaling in the squeezed limit,
	rather than scaling
	of the amplitude
	for nearly equilateral momenta.
	Later, Tzavara \& van Tent
	studied the $k_t$-dependence and the
	scaling behavior of~\eqref{eq:b-scaling}
	in a two-field model
	using a Green's function formalism~\cite{Tzavara:2012qq}.
	In this \emph{Letter},
	we give expressions using the
	separate universe formalism, in both its
	`variational' and `transport'
	versions,
	which are
	valid for any number of scalar fields.
These expressions are only valid when super-horizon evolution of curvature perturbations has ceased, in other words, when the adiabatic limit has been reached. Otherwise, if isocurvature perturbations persist at the end of inflation, it is necessary to augment the model by providing a prescription for their subsequent evolution. Our formulae apply when it is sufficient to evaluate the curvature perturbation at or before the end of the inflationary phase.
	
	\para{The scalar spectral index}%
	Sasaki \& Stewart
	used the separate universe
	method to give
	an expression for $n_s$~\cite{Sasaki:1995aw}.
	More recently an alternative (but equivalent)
	prescription was given in Ref.~\cite{Dias:2011xy},
	which we follow here \footnote{See Ref.~\cite{Byrnes:2006fr}
		for an earlier application of a similar method.}.
	According to the separate universe picture,
	the power spectrum 
	evaluated at time $t$
	can be written as
	\begin{equation}
		\P{}_\zeta(k)|_t
		= N_\alpha N_\beta \Sigma_{\alpha\beta}(k)|_{t_0} ,
		\label{eq:power-spec-sep-univ}
	\end{equation}
	where $N_\alpha \equiv \partial N(t, t_0) / \partial \phi_\alpha(t_0)$,
	and $N$ measures the number of efolds elapsed from a spatially flat
	slice at time $t_0$ to a uniform density hypersurface at later time $t$;
	indices $\alpha$, $\beta$, \ldots,
	label the different species of scalar field;
	and $\Sigma_{\alpha\beta}$ defines the
	two-point function of scalar field fluctuations,
	\begin{equation}
		\langle
			\delta \phi_\alpha(\vect{k}_1)
			\delta \phi_\beta(\vect{k}_2)
		\rangle_{t_0}
		=
		(2\pi)^3 \delta(\vect{k}_1 + \vect{k}_2)
		\frac{\Sigma_{\alpha\beta}|_{t_0}}{2k^3} ,
	\end{equation}
	with $k=|\vect{k}_1|=|\vect{k}_2|$.
	The
	time $t_0$ can be chosen to coincide with the horizon
	crossing time of $k$, which we will denote with a subscript `$k$'. 
	
	The spectral index
	can be obtained from~\eqref{eq:power-spec-sep-univ}
	provided we know 
	$\Sigma_{\alpha\beta}$ to next-order in slow-roll,
	which was first obtained by
	Nakamura \& Stewart~\cite{Nakamura:1996da, *Avgoustidis:2011em}.
	At this order and for light, canonically-normalized fields, we find
	\begin{equation}
		\Sigma_{\alpha\beta}
		=
		H_\ast^2 \left(
			\delta_{\alpha\beta}
			+ 2 r^\ast_{\alpha\beta}
			- 2 u^\ast_{\alpha\beta} \ln (-k_\ast \tau)
			- 2 M^\ast_{\alpha\beta} \ln \frac{2k}{k_\ast}
		\right) ,
		\label{eq:power-spec-nlo}
	\end{equation}
	where `$\ast$' denotes
	evaluation at the horizon-crossing time
	for an arbitrary scale $k_\ast$;
	the conformal time $\tau$ satisfies
	$\d t = a(t) \, \d \tau$;
	$r_{\alpha\beta}$ is a constant;
	$u_{\alpha\beta} = - m_{\alpha\beta}/3H^2$ is
	a rescaled mass matrix for the fluctuations,
	equivalent to the expansion tensor
	of the inflationary flow field~\cite{Seery:2012vj};
	$M_{\alpha\beta} \equiv  \varepsilon \delta_{\alpha\beta}
	+ u_{\alpha\beta}$, and  $\varepsilon \equiv  - \dot{H}/H^2$.

    The $k$-dependence can be extracted from the
    coefficient of
    $\ln k/k_\ast$, provided we know how to
    choose $k_\ast$.
    To do so, note that
	the $\ln(-k_\ast \tau)$ term
	is the lowest power in a series expansion
	which describes
	the time dependence of the fluctuations.
	Since we wish to estimate the scale dependence at a
	fixed time $t_0$ we can choose the arbitrary scale
	so that $k_\ast \tau =-1$,
	making all powers of $\ln(-k_\ast \tau)$ negligible.
	With this choice the evaluation point of
	$H$, $r_{\alpha\beta}$, $u_{\alpha\beta}$ and
	$M_{\alpha\beta}$ becomes coincident with $t_0$
	and
	the spectral index simplifies to
	\begin{equation}
		n_s - 1
		=
		- 2
		\frac{N_\alpha N_\beta M_{\alpha\beta}|_{k}}
			 {N_\lambda N_\lambda} .
		\label{eq:ns-simple}
	\end{equation}

	\para{Squeezed bispectrum}%
	A similar approach can be used to obtain the spectral
	index $\nsq$ in \eqref{eq:b-scaling}.
	We require a next-order expression for the
	bispectrum,
	in analogy with~\eqref{eq:power-spec-nlo}.
	The requisite
	corrections
	are known for general single-field
	models~\cite{Chen:2006nt, *Burrage:2011hd, *Ribeiro:2012ar},
	and were recently obtained for multiple canonically-normalized
	fields in Ref.~\cite{Dias:2012qy}.
	In the squeezed limit they can be written
	\begin{widetext}
	\begin{equation}
		\langle
			\delta\phi_\alpha(\vectkL)
			\delta\phi_\beta(\vectkS)
			\delta\phi_\gamma(\vectkS')
		\rangle
		=
		(2\pi)^3
		\delta(\vectkL + \vectkS + \vectkS')
		B_{\alpha\beta\gamma}
		\overset{\kL \rightarrow 0}{\longrightarrow}
		(2\pi)^3
		\delta(\vectkS + \vectkS')
		\frac{1}{4\kL^3 \kS^3}
		\Big(
			b_{\alpha\mid\beta\gamma} 
			+ 
			\Or\bigg( \frac{\kL^2}{\kS^2} \bigg)
		\Big) .
		\label{eq:squeezed-bispectrum}
	\end{equation}
    Corrections of order 	$\sim \kL^2/\kS^2$ may be relevant 	in the `not-so squeezed' limit
	where the hierarchy $\kL/\kS$ is modest~\cite{Creminelli:2011rh},
	but in the present case we expect their effect to be
	negligible.
	The coefficient $b_{\alpha\mid\beta\gamma}$
	is symmetric under interchange of the indices $\beta$ and $\gamma$, but not
	necessarily under other permutations.
	In what follows it is helpful to break it into
	terms of lowest-order
	and next-order in slow-roll,
	which we label `$\LO$' and `$\NLO$'.
	Using the 
	results of Ref.~\cite{Dias:2012qy},
	we find
	\begin{subequations}
	\begin{align}
		b^{\LO}_{\alpha\mid\beta\gamma}
		& =
		- H_\ast^4 \frac{\dot{\phi}_\alpha}{H_\ast} \delta_{\beta\gamma}
		+ \cdots,
		\label{eq:b-lo}
		\\
		b^{\NLO}_{\alpha\mid\beta\gamma}
		& =
		\left(
			- 2 H_\ast^4 u_{\alpha\beta\gamma}
			- u^{\ast}_{\alpha\lambda} b^{\LO}_{\lambda\mid\beta\gamma}
			- u^{\ast}_{\beta\lambda} b^{\LO}_{\alpha\mid\lambda\gamma}
			- u^{\ast}_{\gamma\lambda} b^{\LO}_{\alpha\mid\beta\lambda}
		\right)
		\ln(-k_\ast \tau)
		- 2 M^\ast_{\alpha\lambda} b_{\lambda\mid\beta\gamma}^{\LO}
		\ln \frac{2\kL}{k_\ast}
		+ \cdots
		\label{eq:b-nlo}
	\end{align}
	\end{subequations}
	\end{widetext}
	The omitted terms are not logarithmically enhanced,
	or are proportional to
	$\ln(\kS/k_\ast)$
	or $\ln(k_t / k_\ast)$.
	In~\eqref{eq:b-lo} and~\eqref{eq:b-nlo} 
	we have chosen $k_\ast \sim k_t$, making these
	contributions negligible
	Therefore, if we intend to estimate this
	expression at some fixed time, only
	$\ln (\kL/k_\ast)$
	can generate dangerously large contributions.
	In analogy with the $\ln(-k_\ast \tau)$ term
	in~\eqref{eq:power-spec-nlo},
    this is the lowest term in a series expansion,
    and, if uncontrolled,
    the series will diverge
    when $\kL / k_\ast \approx \kL / \kS \rightarrow 0$.
    It is for this reason that there is no analytic
    formula for the squeezed limit of the bispectrum in
    a multiple-field model.

    Ref.~\cite{Dias:2012qy} developed a method based on
    the dynamical renormalization group
    to control the series of $\ln(-k_\ast \tau)$ terms
    in the limit $\tau \rightarrow 0$.
	It is not yet clear whether a similar scheme could be used to
	deal with the series expansion of $\ln (\kL/\kS)$.
	For this reason our results
	are untrustworthy for $\kL \ll \kS$.
	Our approach may be reliable up to a
	hierarchy $|\ln (\kL/\kS)|$ of order a few.
	
	Using
	 the separate universe formula, the bispectrum $B_\zeta$
	can be written
	(for an arbitrary momentum triangle)
	\begin{widetext}
	\begin{equation}
	\begin{split}
		B_\zeta(k_1, k_2, k_3)|_t
		= \mbox{}
		&
			N_\alpha N_\beta N_\gamma B_{\alpha\beta\gamma}(k_1, k_2, k_3)|_{t_0}
		\\ &
			+ N_{\alpha\beta} N_\gamma N_\delta
			\bigg(
			\frac{\Sigma_{\alpha\gamma}(k_1)}{k_1^3} 
			\frac{\Sigma_{\beta\delta}(k_2)}{k_2^3}
			+ \frac{\Sigma_{\alpha\gamma}(k_1)}{k_1^3}
			 \frac{\Sigma_{\beta\delta}(k_3)}{k_3^3}
			+ \frac{\Sigma_{\alpha\gamma}(k_2)}{k_2^3} 
			\frac{\Sigma_{\beta\delta}(k_3)}{k_3^3}
			\bigg)|_{t_0} ,
		\label{eq:B-separate-universe}
	\end{split}
	\end{equation}
	where $N_{\alpha\beta} \equiv \partial^2 N(t, t_0)/
	\partial \phi_\alpha(t_0) \partial \phi_\beta(t_0)$.
	The squeezed spectral index, $n_{\rm sq}$,
	can be computed using this expression.
    We take $t_0$ to be the horizon crossing time for
    $k_t$  and denote evaluation
	at this time with a subscript `$k_t$'.
	Choosing $k_\ast \sim k_t$,
	the term proportional to $\ln(-k_\ast \tau)$ in~\eqref{eq:b-nlo} vanishes and 
	the scale dependence of
	$B_{\alpha\beta\gamma}$ can be read off.
	In combination with~\eqref{eq:ns-simple}, we obtain
		\begin{equation}
		\nsq =
		-2 \frac{N_\alpha N_\beta N_\gamma (M_{\alpha\lambda} b^{\LO}_{\lambda\mid\beta\gamma})|_{k_t}
				 + N_{\alpha\beta} N_\gamma N_\delta
				 \big(
				 	M_{\alpha\lambda} \Sigma_{\lambda\gamma}
				 	+ M_{\gamma\lambda} \Sigma_{\alpha\lambda}
				 \big)|_{k_t}
				 \Sigma_{\beta\delta}|_{k_t}}
				{N_\lambda N_\mu N_\nu b^{\LO}_{\lambda\mid\mu\nu}|_{k_t}
				 + 2 N_{\lambda\mu} N_\nu N_\pi
				 (\Sigma_{\lambda\nu} \Sigma_{\mu\pi})|_{k_t}} .
		\label{eq:nb}
	\end{equation}
	\end{widetext}

			Whenever the bispectrum in a multiple-field inflationary
	model is large enough to be observable, the
	`nonlinear' terms in the second line of~\eqref{eq:B-separate-universe}
	dominate the `intrinsic' term
	in the first line~\cite{Lyth:2005qj,Vernizzi:2006ve}.
	In this limit, 
	since $\Sigma_{\alpha\beta}|_{k_t} \approx H^2_{k_t} \delta_{\alpha\beta}$,
	we can write
	$\nsq$ in the simpler form
	\begin{equation}
		\nsq \approx
		- 2 \frac{N_{\alpha\beta} N_\gamma N_\beta M_{\alpha\gamma}|_{k_t}}
				 {N_{\lambda\mu} N_\lambda N_\mu} \ ,
				 \label{reducing}
	\end{equation}
	and under the same assumptions we conclude that
	\begin{equation}
		\nbias \approx
		- 2 \frac{N_{\alpha\beta} N_\gamma N_\beta M_{\alpha\gamma}|_{k_t}}
				 {N_{\lambda\mu} N_\lambda N_\mu}
		+ 2 \frac{N_\alpha N_\beta M_{\alpha\beta}|_{k_t}}
				 {N_\lambda N_\lambda} .
	\end{equation}
	This constitutes one of the principal results of this \emph{Letter}.
	
%
%

	\para{Transport equations}%
	Eqs.~\eqref{eq:ns-simple} and~\eqref{eq:nb} are framed in terms of the
	`variational' formulation of the separate universe
	method, which involves `variational'
	derivatives such as $N_\alpha$.
	A successful numerical implementation
	requires sufficient resolution to extract
	the variation of $N$ 
	against a background of numerical noise accumulated during the
	integration from $t_0$ to $t$.
	An alternative approach to solving~\eqref{eq:power-spec-sep-univ}
	and~\eqref{eq:B-separate-universe} consists in integrating
	the correlators of field fluctuations 
	using 
	Jacobi evolution equations \cite{Seery:2012vj}.
	Mathematically,
    both methods are equivalent,
	but the Jacobi approach is
	simpler because it
	is more tolerant to numerical noise.
	For details on this formalism we refer to
	Refs.~\cite{Mulryne:2009kh,Mulryne:2010rp,
	Seery:2012vj,Anderson:2012em,Elliston:2012ab,Mulryne:2013uka}.

	To write a Jacobi-type equation which defines the time evolution of the
	coefficient matrix $b_{\alpha\mid\beta\gamma}$ in~\eqref{eq:squeezed-bispectrum},
	we define
	$u_{\alpha\beta\gamma} \equiv \partial u_{\alpha\beta} / \partial \phi_\gamma$.
	Using the prescription of Ref.~\cite{Dias:2012qy}, it follows that	\begin{equation}
		-\frac{\d b_{\alpha\mid\beta\gamma}}{\d \ln \tau}
		=
		u_{\alpha\lambda} b_{\lambda\mid\beta\gamma}
		+
		u_{\beta\lambda} b_{\alpha\mid\lambda\gamma}
		+
		u_{\gamma\lambda} b_{\alpha\mid\beta\lambda}
		+
		2 u_{\alpha\lambda\mu}
		\Sigma_{\lambda\beta}
		\Sigma_{\mu\gamma} .
		\label{eq:b-rge}
	\end{equation}
	Note that we should regard
	one of the two-point functions $\Sigma_{\alpha\beta}$ 
	to be evaluated at $\kL$ and the other at $\kS$.
	A suitable initial condition at horizon-crossing
	is provided
	by the lowest-order value~\eqref{eq:b-lo},
	after setting $k_\ast \sim k_t$.

    To compute the scale dependence in the squeezed limit,
	we adjust $\kL$ while keeping $\kS$ fixed.
	Defining $\db_{\alpha\mid\beta\gamma} \equiv  \d b_{\alpha\mid\beta\gamma} / \d \ln \kL$, it follows that
	\begin{equation}
		\begin{split}
		-\frac{\d \db_{\alpha\mid\beta\gamma}}{\d \ln \tau}
		& =
		u_{\alpha\lambda} \db_{\lambda\mid\beta\gamma}
		+
		u_{\beta\lambda} \db_{\alpha\mid\lambda\gamma}
		+
		u_{\gamma\lambda} \db_{\alpha\mid\beta\lambda}\\
		& \hspace{5mm} +
		u_{\alpha\lambda\mu}
		n_{\lambda\beta}
		\Sigma_{\mu\gamma}
		+
		u_{\alpha\lambda\mu}
		\Sigma_{\lambda\beta}
		n_{\mu\gamma} ,
		\end{split}
		\label{eq:db-rge}
	\end{equation}
	where $n_{\alpha\beta}\equiv \d\Sigma_{\alpha\beta}/\d\ln k$. The initial condition 
	 is given by $\db_{\alpha\mid\beta\gamma} = - 2 M_{\alpha\lambda} b^{\LO}_{\lambda\mid\beta\gamma}$
	near horizon-exit for $k_t$.

    Likewise $\Sigma_{\alpha\beta}$
	and $n_{\alpha\beta}$ satisfy \cite{Dias:2011xy}
	\begin{subequations}
	\begin{align}
		\frac{\d \Sigma_{\alpha\beta}}{\d N}
		& =
			u_{\alpha\lambda} \Sigma_{\lambda \beta}
			+ u_{\beta\lambda} \Sigma_{\alpha \lambda} , \\
		\frac{\d n_{\alpha\beta}}{\d N}
		& =
			u_{\alpha\lambda} n_{\lambda \beta}
			+ u_{\beta\lambda} n_{\alpha \lambda} ,
	\end{align}
	\end{subequations}
	with respective initial conditions 
	$\Sigma_{\alpha\beta} = H_\ast^2 \delta_{\alpha\beta}$
	and $n_{\alpha\beta} = - M_{\alpha\lambda} \Sigma_{\lambda\beta} -
	M_{\beta\lambda} \Sigma_{\alpha\lambda}$
	near the horizon-crossing time for $k_t$, 
	supplied by~\eqref{eq:power-spec-nlo}.

    To relate
    $n_{\alpha\beta}$ and $n_{\alpha\mid\beta\gamma}$
    to the spectral indices
    $n_{s}$ and $\nsq$
    requires the gauge transformation
    to $\zeta$.
    We label these coefficients
    $N^t_\alpha$ because they are
    obtained by computing a perturbation
    in the e-folding number $N$,
    but the label $t$ emphasizes that they involve
	only local quantities at $t$.
    Explicit expressions for these gauge transformations are tabulated
	in Anderson et al.~\cite{Anderson:2012em}.
	The spectral index of the bias can
	be written
%
	\begin{equation} 
		\nbias =
			\frac{N^t_\alpha N^t_\beta N^t_\gamma
				  \db_{\alpha\mid\beta\gamma}
				  + 2 N^t_\alpha N^t_\beta N^t_{\gamma\delta}
				  n_{\alpha\gamma} \Sigma_{\beta\delta}}
				 {N^t_\alpha N^t_\beta N^t_\gamma
				  b_{\alpha\mid\beta\gamma}
				  + 2 N^t_\alpha N^t_\beta N^t_{\gamma\delta}
				  \Sigma_{\alpha\gamma} \Sigma_{\beta\delta}}
			 -  \frac{N^t_\alpha N^t_\beta n_{\alpha\beta}}
			 {N^t_\lambda N^t_\mu \Sigma_{\lambda \mu}} .
	\end{equation}
	Focusing on multiple-field models with a bispectrum large enough to be detectable,
	in analogy with Eq.~\eqref{reducing}, this expression reduces to
	\begin{equation} 
		\nbias =
			\frac{N^t_\alpha N^t_\beta N^t_{\gamma\delta}
				  n_{\alpha\gamma}}
				 {N^t_\alpha N^t_\beta N^t_{\gamma\delta}
				  \Sigma_{\alpha\gamma}}
			 -  \frac{N^t_\alpha N^t_\beta n_{\alpha\beta}}
			 {N^t_\lambda N^t_\mu \Sigma_{\lambda \mu}} .
	\end{equation}

\begin{figure}
\hfill
		\subfloat[][Double quadratic model\label{DQ}]{
		\includegraphics[scale=0.17]{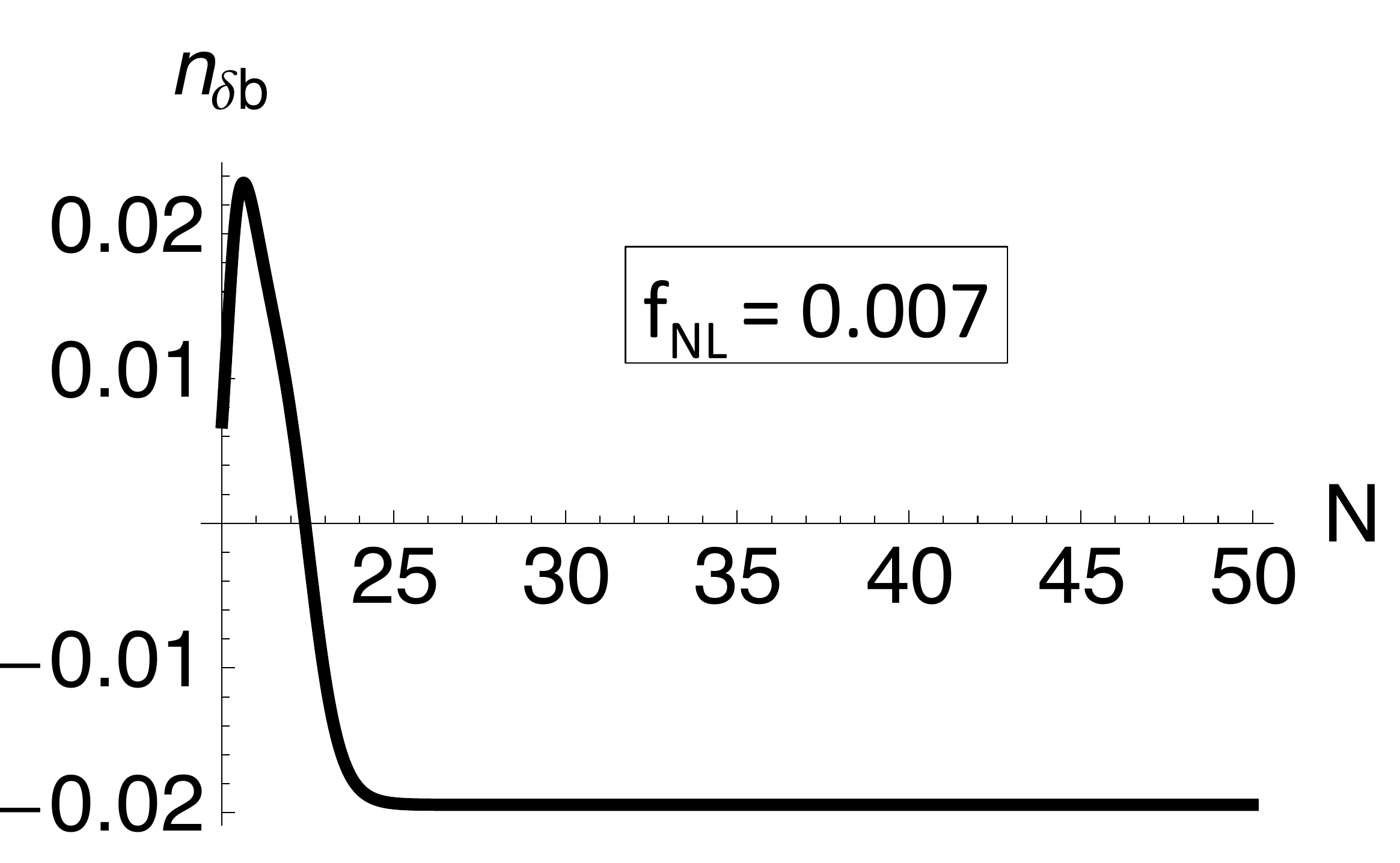}
	}
	\hfill
		\subfloat[][$\phi^4$/axion model\label{Axion}]{
			\includegraphics[scale=0.18]{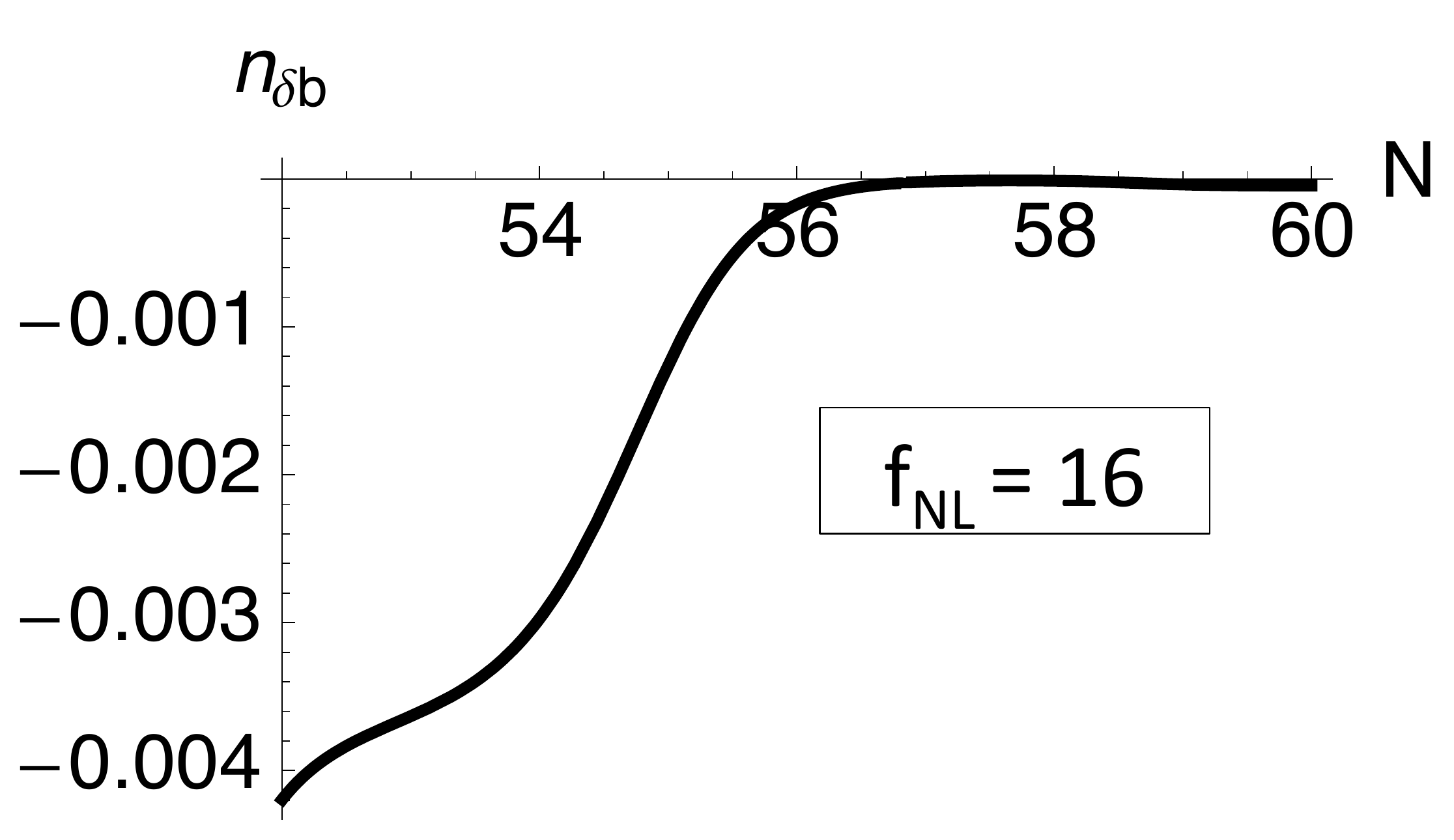}
		}
		
		\caption{Evolution of $\nbias$ some efolds before the end of inflation in two different models. The values of $f_{\rm NL}$ shown are estimated at the end of inflation.}
		\label{fig:ndeltab}
\end{figure}

	As an example, Fig.~\ref{fig:ndeltab} 
	depicts the evolution of
	$\nbias$
	in double quadratic inflation~\cite{Silk:1986vc,Polarski:1994rz,
	GarciaBellido:1995qq,Langlois:1999dw}
	and in the $\phi^4$-plus-axion model studied by
	Elliston~et~al.~\cite{Elliston:2012wm}.
	To generate a significant
	$\nbias$, the bispectrum must scale
	differently to the power spectrum.
	If $|\fNL|$ is small 
	this is typical, as in
	Fig.~\ref{DQ}.
	Alternatively, if $|\fNL|$ is ${\mathcal O}(1)$ or larger the scaling is more similar, as in Fig.~\ref{Axion}.
    Where the spectral index of the bias is large
    we find that it is often negative,
    giving the squeezed limit a red tilt.
    This is the opposite of quasi-single-field inflation
    (QSFI),
    where $\nbias > 0$.
    This can be understood heuristically.
    QSFI contains massive modes, which mediate
    finite-range forces and tend to soften long-range
    correlations.
    In comparison, models with multiple active fields
    tend to add tachyons to the spectrum which
    mediate long-range forces and therefore enhance
    correlations.
    
%

	\para{Conclusions}%
	The halo bias is known to inherit a scale dependence
	from the underlying inflationary model.
	It is therefore an important observable
	which is likely to be measured by
	future surveys of the
	galaxy distribution.
	In this \emph{Letter} we focus on its
	scaling,
	for which we do not require detailed
	knowledge of the amplitude $A_{\mathrm{NL}}$.
	We provide a formula to compute $\nbias$ for any model of multiple-field inflation.
	To obtain a significant effect, the squeezed limit
	(characterized by the spectral index $\nsq$)
	should scale differently
	to the power spectrum
	(characterized by 
	$n_s-1$).
	This can be measured most easily when
	$\nbias < 0$, making
	$k^{-2 + \nbias}$ diverge
	more strongly in the
	limit $k \rightarrow 0$.
	Therefore, a slightly red-tilted bias
    will be easier
	to constrain than a blue-tilted one.
	
	The principal drawback of our method 
	is the use of a spectral index
	to parametrize the behavior of the squeezed bispectrum.
	For a sufficiently large variation of $\kL$
	the bispectrum may have a shape which cannot be
	approximated by a power law.
	Nevertheless, our formula should give
	a good approximation provided the
	hierarchy $\ln (\kL/\kS)$ does not exceed $\sim$ a few.
	Near-future surveys such as DES may probe a hierarchy of order
	$\ln (\kL/\kS) \sim -2$,
	but surveys arriving in the medium- to
	long-term, such as Euclid, may probe
	$\ln (\kL/\kS) \sim -8$.
	Small-scale observables such as $\mu$-distortions
	may even probe $\ln (\kL/\kS) \sim -19$.
	For such cases
	a more precise description of the bispectrum will be
	required, perhaps using numerical
	methods.

	We thank Chris Byrnes, Cora Dvorkin, Simone Ferraro, Jonathan Frazer and Daan Meerburg for discussions.
	DS and MD acknowledge support from 
	STFC [grant number ST/ I000976/1]
	and the Leverhulme Trust.
	DS acknowledges that
	this material is based upon work supported in part by the
	NSF under Grant No. 1066293 and
	the hospitality of the Aspen Center for Physics.
	The research leading to these results has received funding from
	the 
	ERC under the European Union's
	Seventh Framework Programme (FP/2007--2013)/ERC Grant
	Agreement No. [308082].

\bibliography{references}

\end{document}